\documentclass[sigconf,screen]{acmart}

\AtBeginDocument{%
  \providecommand\BibTeX{{%
    \normalfont B\kern-0.5em{\scshape i\kern-0.25em b}\kern-0.8em\TeX}}}

\copyrightyear{2023}
\acmYear{2023}
\setcopyright{rightsretained}
\acmConference[HOPC '23]{Proceedings of the 2023 ACM Workshop on Highlights of Parallel Computing }{June 16, 2023}{Orlando, FL, USA} \acmBooktitle{Proceedings of the 2023 ACM Workshop on Highlights of Parallel Computing (HOPC '23), June 16, 2023, Orlando, FL, USA} \acmDOI{10.1145/3597635.3598022}
\acmISBN{979-8-4007-0218-1/23/06}

\usepackage[normalem]{ulem}

\usepackage{tikz}

\usepackage[flushleft]{threeparttable}
\usepackage{multirow}
\usepackage{multicol}
\usepackage{makecell}

\usepackage[normalem]{ulem}

\usepackage{amsmath,amsfonts}
\usepackage{graphicx}
\usepackage{xcolor}

\usepackage{siunitx}

\usepackage{xcolor}
\usepackage{xspace}
\usepackage{soul}

\usepackage{amsmath}
\usepackage{algorithm}
\usepackage{algpseudocode}
\usepackage{booktabs}
\usepackage{listings}
\usepackage{graphicx}
\usepackage{epstopdf}
\usepackage{subfigure}
\usepackage{multirow}
\usepackage{ulem}
\usepackage{rotating}
\usepackage{enumitem}

\usepackage{MnSymbol}
\usepackage{stmaryrd}

\usepackage{color, colortbl}
\definecolor{Gray}{gray}{0.9}

\begin{document}

\title{CommonGraph: Graph Analytics on Evolving Data (Abstract)} 


\author{Mahbod Afarin}
\authornote{Both authors contributed equally to this research.}
\email{mafar001@ucr.edu}
\affiliation{%
  \institution{CSE Department, UC Riverside}
  \country{USA}
}

\author{Chao Gao}
\authornotemark[1]
\email{cgao037@ucr.edu}
\affiliation{%
  \institution{CSE Department, UC Riverside}
  \country{USA}
}

\author{Shafiur Rahman}
\email{mrahm008@ucr.edu}
\affiliation{%
  \institution{CSE Department, UC Riverside}
  \country{USA}
}

\author{Nael Abu-Ghazaleh}
\email{nael@cs.ucr.edu}
\affiliation{%
  \institution{CSE Department, UC Riverside}
  \country{USA}
}

\author{Rajiv Gupta}
\email{rajivg@ucr.edu}
\affiliation{%
  \institution{CSE Department, UC Riverside}
  \country{USA}
}

\begin{abstract}
We consider the problem of graph analytics on evolving graphs.  In this scenario, a query typically needs to be applied to different snapshots of the graph over an extended time window. We propose \emph{CommonGraph}, an approach for efficient processing of queries on evolving graphs.  We first observe that edge deletions are significantly more expensive than addition operations.  \emph{CommonGraph} converts all deletions to additions by finding a common graph that exists across all snapshots. After computing the query on this graph, to reach any snapshot, we simply need to add the missing edges and incrementally update the query results. \emph{CommonGraph} also allows sharing of common additions among snapshots that require them, and breaks the sequential dependency inherent in the traditional streaming approach where snapshots are processed in sequence, enabling additional opportunities for parallelism. We incorporate the \emph{CommonGraph} approach by extending the KickStarter streaming framework. \emph{CommonGraph} achieves 1.38$\times$-8.17$\times$ improvement in performance over Kickstarter across multiple benchmarks.
\end{abstract}



\maketitle

\section{Motivation}

Analyses on large graphs are an increasingly important computational workload as graph analytics is employed in many domains to uncover insights by mining high volumes of connected data. Graphs are often dynamic, with edges and vertices being added or removed over time. There are two broad classes of analyses on dynamic graphs: (1) \emph{Streaming graph analytics}: where results of a query are continuously updated as the graph continues to change because updates to it stream in over time. Typically incremental algorithms are employed to update query results in response to graph changes~\cite{kickstarter,jetstream,tripoline,yin2022glign}; and (2) \emph{Evolving graph analytics}: where multiple snapshots of the graph are available and an evolving graph query seeks to \emph{track a graph property over a time window} by computing its value for different snapshots within the time window. Such evaluation is computationally expensive. A typical approach for the evolving scenario is to start with query evaluation the earliest snapshot and then use streaming to incrementally evaluate the query on snapshots in sequence. This approach has a number of drawbacks. First, for many algorithms the cost of edge deletions is very high. Second, the solution moves between snapshots in sequence which limits opportunities for sharing query evaluation work among them.  


When incremental algorithms are used, as we move from one snapshot to the next, the graph for the snapshot is first mutated to obtain the one for the next snapshot and then the incremental algorithm is used to update query results. A primary observation that motivates \emph{CommonGraph} is that when the system has to handle both deletions and additions, the cost of incremental computation that handles deletions is significantly higher than that for additions. Moreover, the cost of graph mutations is also significant. The incremental cost of processing a batch of deletions is nearly 3$\times$ the cost of processing an equal number of additions. Handling deletions is more expensive because the algorithm is more complex for monotonic algorithms and impacts on query results are more widespread across the graph, necessitating significantly more processing. Finally, Kickstarter's cost of graph mutation is greater for deletions than additions. This work is an abstracted version of the CommonGraph paper~\cite{CommonGraph}. 


\section{Solution}

\begin{figure}[!t]
    \centering
    \includegraphics[width=1.0\columnwidth]{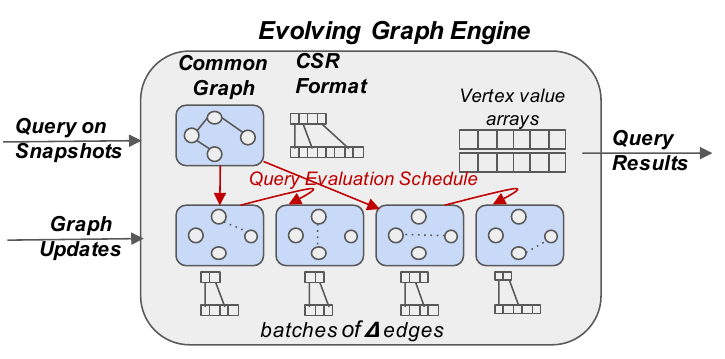}
     \vspace{-0.155in}
    \caption{\emph{CommonGraph} Approach.}
    \label{fig:overview}
        \vspace{-0.2in}
\end{figure}

To address the above problems (i.e., the high incremental cost of deletions and the significant cost of mutation), our system shown in Figure~\ref{fig:overview} based on \emph{CommonGraph} introduces these three complementary techniques. The graph is represented in form of the shared \emph{CommonGraph} and additional batches of edges ($\Delta$ batches) that can be used in conjunction with the \emph{CommonGraph} to realize different snapshots. This representation of the graph and its updates allows different query evaluation schedules (shown as red arrows) to be realized that do not require deletions, incorporate work sharing, and do not require explicit graph mutation. Next, we provide an overview of these three features.

\paragraph*{Converting Deletions to Additions} To overcome the high computational cost of processing edge deletions, we make a key observation: all deletions can be converted to additions by computing a \emph{CommonGraph} that includes only the edges that are common to all the snapshots under consideration. Once query results have been computed on this graph, then results for any snapshot can be computed by \emph{adding the batch of missing edges} to the \emph{CommonGraph} and employing the incremental algorithm to update query results. That is, we can avoid the use of the more expensive incremental algorithm for deletions since deletions become additions if we reverse the order in which the snapshots are processed.

\paragraph*{Work Sharing for a Large Number of Snapshots}
Although the \emph{CommonGraph} achieves work sharing among all the snapshots to process the common graph itself, as the time window grows and the number of snapshots increases, additional opportunities for work sharing of the graph updates among subsets of snapshots arise.  Specifically, any subset of the snapshots may share additional edges in common, and can share work if we stream the additional edges to reach this larger common graph together instead of each query adding them separately to each snapshot.  Assume that we have $n$ snapshots to consider. Our second contribution, the \emph{Triangular Grid} representation, allows systematic exploration and discovery of a $n$ incremental computations that result in query results for all $n$ snapshots while at the same time maximizing the \emph{work sharing}.



\section{Key Results}

We evaluate \emph{CommonGraph} on five benchmarks and five graphs on 50 snapshots. Each snapshot is separated from the next by a batch of 75K edge changes split \emph{evenly} between additions and deletions. The first row for each benchmark in the Table~\ref{tab:time} is baseline \emph{KickStarter}: we start from the initial snapshot and stream in the batches to reach the next snapshot repeatedly. In the second row for each benchmark, we can see the speedup for \emph{CommonGraph}  with Direct-Hop; it outperforms the baseline KickStarter 1.02$\times$-7.91$\times$; even though it processes a higher number of edges compared to KickStarter, all these edges are additions and benefit also from parallelism among additions since they are processed in a single batch.  Moreover, some of the benefits come from avoiding the cost of graph mutation through our graph representation.  Additional speedup is achieved using work sharing, for an overall speedup of 1.38x-8.17x over baseline KickStarter.

\begin{table}[!t]
\captionsetup{justification=centering}
\caption{Average Execution Times in Seconds for KickStarter (KS), and the speedup of CommonGraph Direct Hop (DH) and Work-Sharing (WS)  over KickStarter for 50 Snapshots.}
\label{tab:time}
\centering
\vspace{-0.1in}
{\renewcommand{\arraystretch}{1.2}
\begin{tabular}{|l|l||c|c|c|c|c|} \hline
\multicolumn{1}{|c|}{\textbf{\textsf{G}}} & \textbf{\textsf{Alg.}} & \textsf{\textbf{BFS}} & \textsf{\textbf{SSSP}} & \textsf{\textbf{SSWP}} & \textsf{\textbf{SSNP}} 
& \textsf{\textbf{VT}} 
\\ \hline \hline

\multirow{3}{*}{\textsf{\textbf{LJ}}} &
    \textsc{KS Time} & 3.43s & 3.88s & 3.69s & 3.75s & 5.17s\\ \cline{2-7} \cline{2-7} 
    & \textsc{DH Spe.} & 1.58$\times$ & 1.07$\times$ & 1.23$\times$ & 1.18$\times$ & 1.02$\times$\\ \cline{2-7}
    & \textsc{WS Spe.} & 1.86$\times$ & 1.43$\times$ & 1.38$\times$ & 1.43$\times$ 
    &  1.62$\times$ 
    \\ \hline \hline
\multirow{3}{*}{\textsf{\textbf{DL}}} &
    \textsc{KS Time} & 27.22s & 27.64s & 27.91s & 27.51s & 31.87s\\ \cline{2-7} \cline{2-7} 
    & \textsc{DH Spe.} & 7.09$\times$ & 7.45$\times$ & 7.3$\times$ & 6.7$\times$ & 7.91$\times$\\ \cline{2-7}
    & \textsc{WS Spe.} & 7.17$\times$ & 8.17$\times$ & 7.64$\times$ & 7.21$\times$ & 8.17$\times$ \\
    \hline \hline
\multirow{3}{*}{\textsf{\textbf{Wen}}} &
    \textsc{KS Time} & 4.65s & 4.59s & 4.72s & 4.20s & 2.03s \\ \cline{2-7} \cline{2-7} 
    & \textsc{DH Spe.} & 4.53$\times$ & 1.32$\times$ & 2.73$\times$ & 2.08$\times$ & 3.24$\times$\\ \cline{2-7}
    & \textsc{WS Spe.} & 4.68$\times$ & 2.42$\times$ & 3.31$\times$ & 2.40$\times$ & 3.8$\times$\\
    \hline \hline 
\multirow{3}{*}{\textsf{\textbf{TTW}}} &
    \textsc{KS Time} & 10.91s & 11.73s & 11.32s & 11.31s & 15.30s\\ \cline{2-7} \cline{2-7} 
    & \textsc{DH Spe.} & 3.09$\times$ & 2.36$\times$ & 2.52$\times$ & 1.85$\times$ & 2.85$\times$\\ \cline{2-7}
    & \textsc{WS Spe.} & 3.35$\times$ & 2.94$\times$ & 3.14$\times$ & 2.62$\times$ & 3.42$\times$ \\
    \hline 
\end{tabular}
}
\vspace{-0.20in}
\end{table}

\section{Key Contribution}

The key contributions of our work are as follows:
\begin{itemize}
    \item A new approach to evolving graphs analysis that converts all graph updates to additions over the \emph{CommonGraph}, replaces expensive deletions by additions, and removes dependencies that enable work sharing among the snapshots.
    \item A structure called {\em Triangular Grid} (TG) that exposes work sharing possibilities among the snapshots. 
    \item A graph representation that avoids the need to mutate graphs and enables reuse of edges by snapshots that share them.
    \item We build the \emph{CommonGraph} that exploits the three ideas above to achieve speedups ranging between 1.38$\times$ and 8.17$\times$ over Kickstarter.
\end{itemize}









\balance
\bibliographystyle{ACM-Reference-Format}
\bibliography{refs}

\end{document}